\documentclass[aps, twocolumn, 8pt]{revtex4}

\usepackage{graphicx,amssymb}

\def\NN{\mathbb N}

\def\II{\mathbb I}

\begin{document}
\title{Security proof for QKD systems with threshold detectors}
\author{Toyohiro Tsurumaru}
\affiliation{Mitsubishi Electric Corporation,
Information Technology R\&D Center,\\
5-1-1 Ofuna, Kamakura-shi, Kanagawa,
247-8501, Japan}
\author{Kiyoshi Tamaki}
\affiliation{NTT Basic Research Laboratories, NTT Corporation, 3-1, Morinosato Wakamiya Atsugi-shi, Kanagawa, 243-0198, Japan\\ CREST, JST Agency, 4-1-8 Honcho, Kawaguchi, Saitama, 332-0012, Japan\\}

\begin{abstract}
In this paper, we rigorously prove the intuition that in security proofs for BB84  one may regard an incoming signal to Bob as a qubit state.
From this result, it follows that all security proofs for BB84 based on a virtual qubit entanglement distillation protocol,
which was originally proposed by Lo and Chau [H.-K. Lo and H. F. Chau, Science 283, 2050 (1999)],
and Shor and Preskill [P. W. Shor and J. Preskill, Phys. Rev. Lett. 85, 441 (2000)], are all valid even if Bob's actual apparatus cannot distill a qubit state explicitly.
As a consequence, especially, the well-known result that a higher bit error rate of 20\% can be tolerated for BB84 by using two-way classical communications is still valid even when Bob uses threshold detectors.
Using the same technique, we also prove the security of the Bennett-Brassard-Mermin 1992 (BBM92) protocol where Alice and Bob both use threshold detectors.
\end{abstract}

\maketitle

\section{Introduction}
Quantum key distribution (QKD) is a way to share secret keys between separated parties (Alice and Bob)
with negligibly small leakage of its information to an unauthorized third party,
Eve. The first QKD protocol, BB84, was introduced by Bennett and Brassard in 1984 \cite{BB84},
and its unconditional security was first proven by Mayers \cite{Mayers} in a bit complicated manner.
After the first proof, researchers have tried to prove its security in a simple manner.
Some proofs are based on the entanglement distillation protocol (EDP) idea \cite{LC98,SP00,GL03,GLLP04}, and others rely on uncertainty principle \cite{Mayers,Koashi} or information-theoretic approach \cite{information}.

In EDP-based security proofs, we usually assume implicitly that Bob has a detector
which can discriminate between vacuum, single-photon, and multi-photon states in order to distill a qubit state, while this is not the case for the security proof based on uncertainty principle \cite{Koashi}, i.e., the conventional on-off detectors (threshold detectors) can be used in this case.
On the other hand, the EDP-based security proof can apply to many protocols, including BB84 with two-way classical communications \cite{GL03}, with decoy states \cite{decoy}, for B92 \cite{Tamaki}, and so on \cite{FTL06}, however, the security proof based on uncertainty principle cannot directly apply to these protocols. Thus, it is important to consider from experimental or theoretical viewpoints how to accommodate the use of threshold detectors in EDP-based security proof, or to consider how to apply the uncertainty principle idea to the other protocols. 

In this paper, we first prove unconditional security of BB84 with threshold detectors based on the argument of virtual EDP,
which is valid regardless of one-way or two-way classical communications.
In order to show its security,
instead of assuming photon-number discriminating detectors,
we introduce an explicit {\it squash operator} in the virtual protocol, which transforms Bob's incoming multi-photon state to a qubit state.
Then we suppose that they run a virtual EDP on the obtained qubit pairs in order to extract secret keys.
If one-way classical communications are used in this setup, the secret key rate $R$ from the single photon part is $R=1-H_2(e^{\rm vi}_{\rm bit})-H_2(e^{\rm vi}_{\rm ph})$,
where $e^{\rm vi}_{\rm bit}$ and $e^{\rm vi}_{\rm ph}$ are the phase and the bit error rates in the virtual protocol.
As a consequence of introducing an explicit squash operator,
all of the formulas for key generation rate given in the preceding literatures of EDP-based security proofs are valid with threshold detectors,
even when multi-photon emission is taken into account \cite{GLLP04} or with decoy states \cite{decoy}.
Our formulation also applies to the case of two-way classical communications \cite{GL03}, hence the bit error rate threshold of 20\% or higher is true with threshold detectors as well.

By using the same technique, we also prove the security of the Bennett-Brassard-Mermin 1992 (BBM92) protocol \cite{BBM92}, where Alice and Bob both use threshold detectors (see Fig.\ref{fig:2a}).
In the BBM92 protocol, a third party supplies entangled states to Alice and Bob, and they measure it with the same set of bases as in BB84.
If both the receivers have photon-number discriminating detectors and can reject incoming multi-photon states,
this protocol is theoretically equivalent to BB84.
When threshold detectors are used, however, the security of this protocol is not straightforward,
and we will give the security proof for this scheme in this paper.

The assumptions that we make for theoretical description of BB84 are as follows.
First, it is assumed that Alice's signals are block diagonalized with respect to photon number,
and thus one can treat events having different photon numbers as distinct classical events.
Moreover, we assume that Alice's mixed states in the $z$ basis and the one in the $x$ basis are the same,
i.e., there is no basis information flow from Alice's source.

We also suppose that when Alice emits a multi-photon state, all information regarding that bit is freely leaked to Eve due to the photon-number splitting attack \cite{BLMS00}.
It is proven, however, that we can still generate a secret key as long as Alice's signals contain a sufficiently high ratio of single-photon states \cite{GLLP04}.
This ratio can be well-monitored by the decoy state method \cite{decoy}, resulting in longer distances of communications.
Thus, only single-photon emission part is important, to which we restrict our attention in this paper.

Another assumption we make is that all imperfections of Alice's and Bob's devices, i.e., non-unit quantum efficiency of Bob's detectors, dark counts, miss-alignment, etc., are under Eve's control.
This is the so-called {\it untrusted devise scenario}, and with this hypothesis we are in a situation where Alice's and Bob's devices are all perfect.
In addition, we suppose that Bob's phase modulator acts on multi-photon states as linear operations on tensor product states.
In other words, they transform each photon contained in a signal independently, whether they are in a superposition or not (for more details, see Sec. \ref{eq:Bob_operations}).

Finally, when Bob's two detectors click simultaneously (coincidence count), he assigns a random bit to the corresponding event.

These assumptions are also made in our security proof of BBM92 except that Alice, as well as Bob, plays the role of a receiver.
That is, imperfections of apparatuses are attributed to Eve's attack,
and Alice's and Bob's phase modulators transform their incoming multi-photon states as tensor products.
If a coincident-detection event occurs on either Alice's or Bob's side, he or she manually replaces it by a random bit. We emphasize in the case of BBM92 that we do not put any assumption on incoming signals.

This paper is organized as follows.
We describe and formulate our model for actual QKD systems based on BB84 in Sec. \ref{sec:our_model},
and convert it into a virtual EDP in Sec. \ref{sec:virtual}.
Sec. \ref{sec:BBM92} is devoted to the security proof for BBM92.
Then finally we conclude in Sec. \ref{sec:conclusion}.

\section{Description of our model}\label{sec:our_model}
\begin{figure}[htbp]
\begin{center}
\includegraphics[clip,scale=0.5]{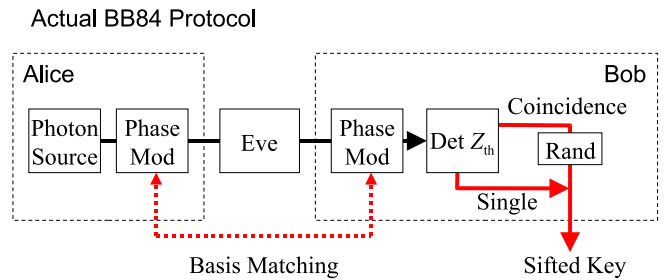}
\end{center}
\caption{(Color online) Schematic of the actual BB84 protocol.
Det $Z_{\rm th}$ denotes Bob's threshold detectors.
When two detectors click simultaneously (coincidence count), Bob assigns a random bit (Rand) to the corresponding event.}
\label{fig:1a}
\end{figure}

In this section we illustrate our setup (Fig.\ref{fig:1a}).
As in usual implementations of the BB84 protocol,
Alice emits out signal pulses whose phases are chosen randomly out of $\{0, \pi, \pi/2, 3\pi/2\}$.
Among them we regard a set of phase choices $\{0, \pi\}$ (respectively, $\{\pi/2, 3\pi/2\}$) as the encodings of bit value $b\in\{0,1\}$ in the $z$ basis
(respectively, in the $x$ basis).
After traveling Eve's regime, the signal pulse is again phase-modulated according to Bob's random basis choice,
and then enters the detection unit consisting of a 50:50 beam splitter followed by two threshold detectors (Det $Z_{\rm th}$), which read out the bit value $b$.
Even when coincident detections occur on both detectors,
Bob does not discard the event and instead assigns a random value for the output $i_B$.

As mentioned in the Introduction, 
the goal of this paper is to rigorously prove the security of QKD even when the receiver (or the receivers) uses threshold detectors which cannot distinguish photon numbers.
Hence throughout the paper, we will always take into account the possibility that states which a receiver obtains contain more than one photon.
To this end, we will below formulate general $N$-photon states and describe how they are transformed by Bob's phase modulations.

\subsection{Symmetry under particle permutations and the formulation of Bob's quantum operations}\label{eq:Bob_operations}
Consisting of identical particles with bosonic statistics,
a state received by Bob is always symmetric under particle permutations \cite{Sakurai}
\footnote{Here we assume that Bob's incoming states are in a single mode.}.
Hence an $N$-particle state in Bob's Hilbert space ${\cal H}_B$ can be expanded with basis
\[
|S^z_{N-b,b}\rangle:=\frac1{\sqrt{N!(N-b)!b!}}\left(|0_z^{N-b}1_z^b\rangle+{\rm permutations}\right),
\]
where $|0_z^{N-b}1_z^b\rangle=|0_z\cdots0_z1_z\cdots1_z\rangle=|0_z\rangle\otimes\cdots\otimes|0_z\rangle\otimes|1_z\rangle\otimes\cdots\otimes|1_z\rangle$,
with $0_z$ and $1_z$ repeating $N-b$ and  $b$ times respectively.
$|S^y_{N-b,b}\rangle_B$ are defined in the $y$ basis similarly (we define $y$ basis and $x$ basis as $|j_y\rangle\equiv (|0_z\rangle+(-1)^j i|1_z\rangle)/\sqrt{2}$ ($j=0,1$) and $|j_x\rangle\equiv (|0_z\rangle+(-1)^j |1_z\rangle)/\sqrt{2}$, respectively).
Thus for example,
$|S^z_{0,2}\rangle=|0_z0_z\rangle$,
whereas $|S^z_{1,1}\rangle=\frac1{\sqrt2}\left(|0_z1_z\rangle+|1_z0_z\rangle\right)$
and $|S^y_{1,2}\rangle=\frac1{\sqrt3}\left(|0_y1_y1_y\rangle+|1_y0_y1_y\rangle+|1_y1_y0_y\rangle\right)$.

Using this basis, quantum non-demolition (QND) measurement of $N$ photon, to be mentioned below,
can be represented by Kraus operators $E^N=\sum_{b=0}^N P(|S^z_{N-b,b}\rangle_B)$,
where $P(|\psi\rangle):=|\psi\rangle\langle\psi|$.

As is usually the case for a linear operator on tensor product states,
or as one typically encounters when adding angular momenta \cite{Sakurai},
Bob's phase modulator acts on these symmetric states independently in a qubit by qubit manner.
For example, bit flip $X$ operates on $|S^z_{3,0}\rangle$ as $X|0_z\rangle\otimes X|0_z\rangle\otimes X|0_z\rangle=: X^{\otimes3}|S^z_{3,0}\rangle=: D(X)|S^z_{3,0}\rangle$,
and similarly the Hamadard gate transforms $|S^z_{1,1}\rangle$ as
$D(H)|S^z_{1,1}\rangle=\frac1{\sqrt2}\left(H|0_z\rangle\otimes H|1_z\rangle+H|1_z\rangle\otimes H|0_z\rangle\right)$, where 
\[
H:=\frac{1}{\sqrt{2}}\left(
\begin{array}{cccc}
1&-1 \\
1&1 
\end{array}
\right)\,
\]
in the $z$ basis.
If one regards qubit operations as rotations of spin-1/2,
these symmetric $N$-photon states correspond to a spin-$N/2$ representation.

In fact these formulations are equivalent to those using creation and annihilation operators. For example, let $a^z_b$ be the annihilation operator corresponding to $|b_z\rangle$, then $|S^z_{N-b,b}\rangle$ corresponds to
\[
|S^z_{N-b,b}\rangle=\frac1{\sqrt{(N-b)!b!}}\left(a^{z\dagger}_0\right)^{N-b}\left(a^{z\dagger}_1\right)^b\left|0\right\rangle
\]
and $D(iX)$ is reproduced by
\[
D(iX)=\exp\left[i\frac{\pi}2\sum_{c,d}a^{z\dagger}_cX_{cd}a^z_d\right].
\]

Photon detection in general corresponds to a projective measurement with respect to photon-number states in the $z$ basis
$\{|S^z_{N,0}\rangle,|S^z_{N-1,1}\rangle,\cdots,|S^z_{0,N}\rangle\}$.
Since Bob's threshold detectors cannot discriminate photon numbers in our case,
it is assumed that they can only distinguish between vacuum $|S^z_{0,0}\rangle$,
single detection events $|S^z_{N,0}\rangle$, $|S^z_{0,N}\rangle$,
and coincident detection events $\{|S^z_{N-1,1}\rangle,\cdots,|S^z_{1,N-1}\rangle\}$.

\section{Virtual protocol}\label{sec:virtual}
\begin{figure}[htbp]
\begin{center}
\includegraphics[clip,scale=0.5]{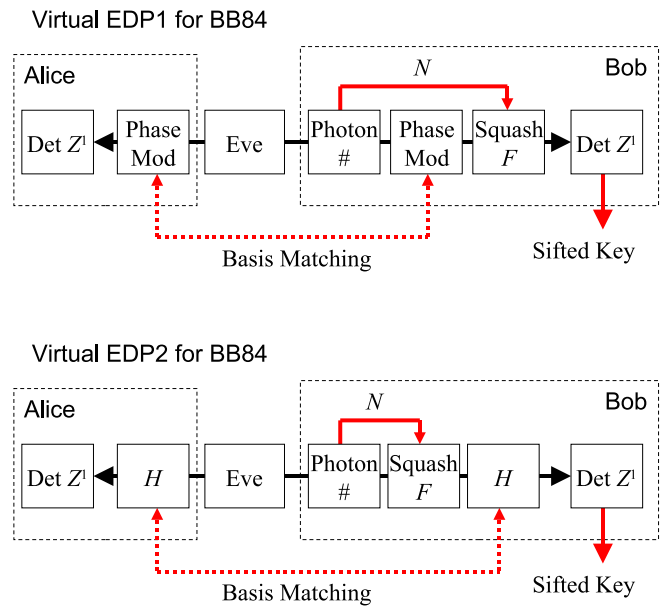}
\end{center}
\caption{(Color online) Schematics of our virtual EDPs for BB84.
In the first protocol (EDP1), Bob's states are projected to a two-dimensional vector space ${\cal H}_D$ by a squash operator $F$.
By using the Hadamard invariance of $F$, we may interchange the order of Bob's phase modulations and $F$, and obtain the second protocol (EDP2).
In EDP2, Alice and Bob both have a qubit space ${\cal H}_A$ or ${\cal H}_D$ with  Hadamard operations acting on them. Thus by performing a virtual EDP in these spaces, they can extract secret keys.
}
\label{fig:1b}
\end{figure}

In this section, in order to prove the security of our QKD, we convert the actual BB84 protocol to an equivalent virtual protocol (Fig. \ref{fig:1b}).
We do this conversion in two steps. First, immediately after Bob's phase modulator,  we introduce an explicit {\it squash operator} $F$, which projects an $N$-particle state received from Eve into a qubit state (see EDP1 in Fig. \ref{fig:1b}). Then by using the invariance of $F$ under the Hadamard transformation $H$, we replace Bob's phase modulator with a Hadamard operation that follows $F$ (EDP2 in Fig. \ref{fig:1b}).
In this EDP2, Alice and Bob obtain a pair of qubit states with Hadamard operators, as if Bob's incoming states were always a single-photon state. Hence we can prove security of our QKD system by following exactly the same argument as in Shor-Preskill \cite{SP00} or in GLLP \cite{GLLP04}. That is, Alice and Bob can perform a virtual EDP to extract secret keys.

First, on Alice's side we assume that, instead of randomly choosing the phase of signal states out of $\{0,\pi/2, \pi, 3\pi/4\}$,
Alice takes the following procedure;
she prepares one of the Bell states
\[
|\Phi^+\rangle_{AE}:=\frac1{\sqrt2}\left(|0_z\rangle_A|0_z\rangle_E+|1_z\rangle_A|1_z\rangle_E\right),
\]
keeps the first one half in ${\cal H}_A$ (reference state), and sends the second  one half in ${\cal H}_E$ (signal state) to Eve.
The converted protocol is still equivalent to the original one,
since she can effectively emit a random bit $i_A\in\{0,1\}$ by measuring the reference state with $z$- and $x$-bases.

On receiving the signal pulse, Eve generates an arbitrary state in Bob's Hilbert state ${\cal H}_B$, which in general may be a superposition of any photon number $N$.
As a result, Alice and Bob end up sharing a state $\rho_{AB}\in {\cal H}_A\otimes{\cal H}_B$. Without sacrificing security, we may simplify the analysis further by assuming that $\rho_{AB}$ is actually given from Eve to Alice and Bob.

Then on Bob' s side, we assume that immediately after receiving a pulse, he performs a QND measurement on the photon number $N$, described by Kraus operators $\{E_N\}_{N\in\NN}$. Since this measurement does not disturb the statistics of Bob's and Eve's data, we can introduce this measurement without loss of generality. To be more precise, Bob's sifted keys are the same regardless of performing QND or not. Moreover, since the measurement outcomes are kept secret from Eve, she cannot behave differently depending on whether he performs the QND or not. In other words, the QND measurement here is introduced only as a convenient tool for proving the security, and Bob need not perform this measurement in the actual protocol.
For the sake of simplicity, we fix the value of $N$ from now on, and sometimes suppress its indices.
Following this measurement,
Bob projects his state to a qubit state using a {\it squash operator} $F$ which converts a state in
his $(N+1)$-dimensional Hilbert space ${\cal H}_B$ to that in a qubit space ${\cal H}_D$.
This consists of Kraus operators
\begin{eqnarray}
\lefteqn{F_{b,b'}:=2^{-(N-1)/2}\times}\label{eq:def_F}\\
&&\left\{\sqrt{N\choose b'}\ |1_y\rangle\langle S^y_{N-b,b}|
+\sqrt{N\choose b}\ |0_y\rangle\langle S^y_{N-b',b'}|\right\},\nonumber
\end{eqnarray}
for all combinations of $0\le b,b'\le N$ satisfying $b-b'\equiv 1\ {\rm mod}\ 4$.
With these operators, an arbitrary state $\rho\in{\cal H}_B$ is converted to
\begin{equation}
F(\rho)=\sum_{b,b'}F_{b,b'}\rho F^\dagger_{b,b'}
\label{eq:rho_F}
\end{equation}
in ${\cal H}_D$.
As shown in Appendix \ref{app:prop_F},
these operators indeed satisfy $\sum_{b,b'} F^\dagger_{b,b'}F_{b,b'}=\II_B$,
and thus form a legitimate quantum operation.

This protocol (EDP1 in Fig. \ref{fig:1b}) can indeed be considered as equivalent to the original one, since, as we shall show in Appendix \ref{app:equiv_z_meas}, the POVM elements in the virtual protocol and in the actual protocol corresponding to Bob's sifted key bit $i_{B}=0,1$ take exactly the same forms. Hence Eve can never distinguish the two protocols.

Moreover, as we shall show in Appendix \ref{sec:Hadamard_invariance}, our squash operation $F$ is in fact invariant under the Hadamard transformation $H$. That is, for an arbitrary input state $\rho\in{\cal H}_B$, we have
\begin{equation}
HF(\rho)H^\dagger=F\left(D(H)\rho D^\dagger(H)\right).
\label{eq:F_invariance}
\end{equation}
In other words, applying the Hadamard transformation in Bob's $(N+1)$-dimensional space ${\cal H}_B$ is equivalent to applying it in the squashed qubit state ${\cal H}_D$. Hence we may assume that Bob's Hadamard transformation and the squash operator $F$ are actually interchanged, as in EDP2 in Fig. \ref{fig:1b}.

Then by supposing that, instead of immediately conducting $z$-basis measurements,
Alice and Bob perform a virtual EDP on a qubit-pair state in their squashed qubit spaces ${\cal H}_A\otimes{\cal H}_D$,
the standard argument for security proofs due to Shor and Preskill \cite{SP00} can be directly applied.
This means that, even when threshold detectors are used, all the previous security proofs for BB84 based on a virtual EDP (e.g., \cite{SP00,GLLP04,GL03,decoy}) are valid, including the ones involving two-way classical communications \cite{GL03} or decoy states \cite{decoy}.

\section{BBM92 protocol with both parties using threshold detectors}\label{sec:BBM92}
By the same arguments as above,
it is straightforward to prove the security of the BBM92 protocol where Alice and Bob both use threshold detectors.

\begin{figure}[htbp]
\begin{center}
\includegraphics[scale=0.36]{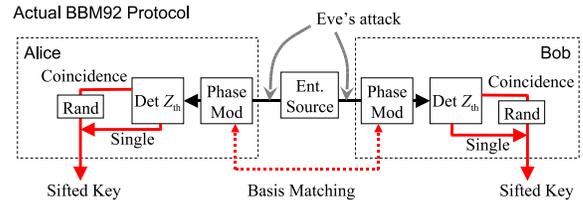}
\end{center}
\caption{(Color online) Schematic of the actual BBM92 protocol.
``Ent. Source" denotes a third party that supplies entangled states to Alice and Bob.
This third party may be malicious in general.}
\label{fig:2a}
\end{figure}

In the actual BBM92 protocol, entangled states are supplied to Alice and Bob from a third party who is not necessarily trusted.
Upon receiving pulses, Alice and Bob modulate and measure them in randomly chosen $x$ or $z$ basis,
and then output sifted key bits by selecting out events where their choices of basis match.
If a coincident-detection event occurs, the receiver assigns a random value to the output bit $i_A$ or $i_B$ (Fig. \ref{fig:2a}).

\begin{figure}[htbp]
\begin{center}
\includegraphics[scale=0.36]{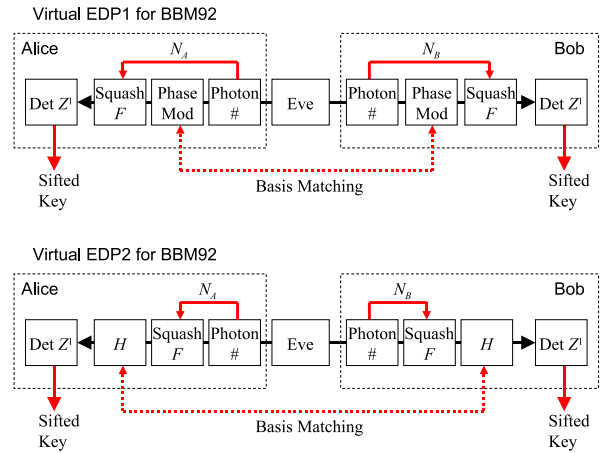}
\end{center}
\caption{(Color online)
Schematics of our virtual EDPs corresponding to BBM92.
In EDP1, Alice and Bob both apply squash operations $F$ and obtain qubit states in ${\cal H}_C$ and ${\cal H}_D$.
By using the Hadamard invariance of $F$, we may interchange the order of phase modulations and $F$'s, and obtain EDP2, where both parties apply the Hadamard gate $H$ on the squashed qubit states ${\cal H}_C$ and ${\cal H}_D$.
}
\label{fig:2b}
\end{figure}

In this case as well, by using the equivalence of POVM elements for $z$-basis measurements (Appendix \ref{app:equiv_z_meas}), this protocol can be equivalently converted into the first virtual protocol (EDP1 in Fig.\ref{fig:2b}), where Alice and Bob apply squash operations $F$ after phase modulations.
Then due to the Hadamard-invariance of $F$ in Eq. (\ref{eq:F_invariance}), we may interchange the order of phase modulations and $F$'s. As a result, we obtain our second virtual protocol (EDP2 in Fig.\ref{fig:2a}), where both parties  apply $H$ on the squashed qubit states ${\cal H}_C$ and ${\cal H}_D$. Here ${\cal H}_C$ denotes Alice's qubit space. In this EDP2, the standard Shor-Preskill-type argument for security proof can be applied to prove the security.

\section{Conclusion}\label{sec:conclusion}
In this paper, we gave a rigorous security proof for BB84, which is valid even when the actual Bob uses threshold detectors.
The key ingredient here was the introduction of an explicit squash operator $F$ in the virtual protocol, which projects Bob's $N$-photon state into a qubit space and still preserves their transformation property under the Hadamard transformation. 
Our results show that all the formulas for key generation rates obtained in previous proofs based on a virtual qubit entanglement distillation protocol are valid even with threshold detectors.
In particular, one can tolerate a higher error rate up to 20\% with two-way classical communications \cite{GL03}.

In addition, by using the same technique, we also proved the security of BBM92 protocol, where Alice and Bob both use threshold detectors.

{\it Note added}.
Recently, similar results regarding construction of squash operators were obtained independently by Beaudry, Moroder and L\"utkenhaus \cite{BML08}. A security proof of BBM92 with threshold detectors was also given independently by Koashi et al. \cite{KAYI08}, although the techniques used there were different from ours.

\ 

\noindent{\large\bf Acknowledgments}

The authors would like to thank T. Moroder, M. Curty, H.-K. Lo and, especially, M. Koashi for enlightening discussions.
This work was supported by the project ``Research and Development on Quantum Cryptography
of the National Institute of Information and Communications Technology,"
as part of Ministry of Internal Affairs and Communications of Japan's program
``R\&D on Quantum Communication Technology."

\appendix
\section{Properties of our squash operator}
In this appendix we discuss and prove some important properties of our squash operator $F$ defined in Eq. (\ref{eq:def_F}).
Without loss of generality, we here consider only Bob's filter, and denote the $(N+1)$-dimensional space of incoming pulses by ${\cal H}_B$ and the target qubit space by ${\cal H}_D$. Note that the arguments proved here can immediately apply when Alice uses $F$ in BBM92 as well.

\subsection{Completeness as a quantum operator}\label{app:prop_F}
First we show that $F$ is a legitimate quantum operation. That is, we demonstrate that
\begin{equation}
F_{\rm sum}=\II_B,\label{eq:F_sum_rel}
\end{equation}
where $\II_B$ is the unit operator in ${\cal H}_B$, and $F_{\rm sum}$ is defined as
\[F_{\rm sum}:=\sum_{b,b'}F_{b,b'}^\dagger F_{b,b'}.\]
To this end, we work in a basis $\{|S^y_{N,0}\rangle,\dots,|S^y_{0,N}\rangle\}$, and prove $f_{b,b'}=\delta_{b,b'}$ for
\[
f_{b,b'}:=\langle S^y_{N-b,b}|F_{\rm sum}|S^y_{N-b',b'}\rangle.
\]

First, it is obvious from the definition of $F$ in (\ref{eq:def_F}) that $f_{b,b'}=0$ for $b\ne b'$.
On the other hand, if $b=b'$, a simple calculation shows that for a fixed value of $b$,
\[
f_{b,b}=2^{-(N-1)}\sum_{b-c\equiv\pm1}{N\choose c},\\
\]
where the sum is over all values of $c$ satisfying $b-c\equiv\pm1\ {\rm mod}\ 4$.
For $b$ even, this equals
\[
f_{b,b}=2^{-(N-1)}\sum_{c:{\rm odd}}{N\choose c}=1,
\]
as anticipated. The case of odd $b$ can be shown similarly.
This completes the proof.

\subsection{Equivalence of $z$ measurements in the actual and the virtual protocols}\label{app:equiv_z_meas}
Next we show that Bob's $z$ measurement in the virtual protocol affects the entire quantum system in exactly the same manner as in the actual one.
That is to say, with the POVM corresponding to a sifted key bit $i_{B}=0,1$ defined as
\begin{eqnarray*}
P^{\rm vi}_{i_{B}}&=&\sum_{c,c'} F^\dagger_{c,c'}P\left(\left|\left(i_B\right)_z\right\rangle\right)F_{c,c'},\nonumber\\
P^{\rm ac}_{i_{B}}&=&P(|S^z_{(1-i_{B})N,i_{B}N}\rangle)+\frac12\sum_{c=1}^{N-1}P(|S^z_{N-c,c}\rangle)\label{eq:eqaulity_ac_vi}
\end{eqnarray*}
respectively for the virtual and the actual protocols, we shall show that $P^{\rm vi}_{i_{B}}=P^{\rm ac}_{i_{B}}$.
If one notes that $P^{\rm vi}_0+P^{\rm vi}_1=\II_B$ holds from Eq.(\ref{eq:F_sum_rel}),
it is rather convenient to consider a POVM element corresponding to the $Z$ operator
\begin{eqnarray*}
P_Z^{\rm ac}&=&P(|S^z_{N,0}\rangle)-P(|S^z_{0,N}\rangle),\\
P_Z^{\rm vi}&=&\sum_{b,b'}F^\dagger_{b,b'}ZF_{b,b'}.
\end{eqnarray*}
and show that
\begin{equation}
P_Z^{\rm ac}=P_Z^{\rm vi}.
\label{eq:PZ_ac_vc_equal}
\end{equation}
Note that $P^{\rm vi}_{i_{B}}=P^{\rm ac}_{i_{B}}$ can be shown from (\ref{eq:PZ_ac_vc_equal}) and by using the relation $P_Z^{\rm vi}=P_0^{\rm vi}-P_1^{\rm vi}$
and $P^{\rm vi}_0+P^{\rm vi}_1=\II_B$, and similar relations for the actual protocol.

Now, from the definition of $F$ in Eq. (\ref{eq:def_F}),
\begin{eqnarray*}
\lefteqn{P_Z^{\rm vi}=2^{-N+1}\sum_{b,b'}\sqrt{{N\choose b}{N\choose b'}}\times}\\
&&\left[|S^y_{N-b,b}\rangle\langle S^y_{N-b',b'}|+|S^y_{N-b',b'}\rangle\langle S^y_{N-b,b}|\right],
\end{eqnarray*}
where the sum is over all $0\le b,b'\le N$ satisfying $b-b'\equiv1\ {\rm mod}\ 4.$
This can be rewritten further as
\begin{equation}
P_Z^{\rm vi}=2^{-N+1}\sum_{b,b'}\sqrt{{N\choose b}{N\choose b'}}|S^y_{N-b,b}\rangle\langle S^y_{N-b',b'}|
\label{eq:PZvi1}
\end{equation}
where the sum is over all $b,b'$ with $b-b'\equiv1\ {\rm mod}\ 2$.

Next, expanding $|S^z_{N-b,b}\rangle$ with $|S^y_{N-b,b}\rangle$ gives
\begin{eqnarray*}
|S^z_{N,0}\rangle&=&2^{-N/2}\sum_{b=0}^N\sqrt{N\choose b}|S^y_{N-b,b}\rangle,\\
|S^z_{0,N}\rangle&=&2^{-N/2}(-i)^N\sum_{b=0}^N\sqrt{N\choose b}(-1)^b|S^y_{N-b,b}\rangle,
\end{eqnarray*}
and from these relations we obtain
\begin{equation}
P_Z^{\rm ac}=2^{-N+1}\sum_{b,b'}\sqrt{{N\choose b}{N\choose b'}}|S^y_{N-b,b}\rangle\langle S^y_{N-b',b'}|,
\label{eq:PZac1}
\end{equation}
where the sum is again over all $0\le b,b'\le N$ satisfying $b-b'\equiv1\ {\rm mod}\ 2$.
Eqs. (\ref{eq:PZvi1}) and (\ref{eq:PZac1}) prove Eq. (\ref{eq:PZ_ac_vc_equal}).
This completes the proof.

\subsection{Invariance under the Hadamard transformation}\label{sec:Hadamard_invariance}
In this subsection we prove in Eq. (\ref{eq:F_invariance}), i.e., the Hadamard-invariance of $F$. Since $|S_{N-b,b}\rangle$ transforms under the Hadamard transformation $H$ as
\[
D(H)|S_{N-b,b}\rangle=\omega^{2b-N}|S_{N-b,b}\rangle,
\]
with $\omega:=\exp(i\pi/4)$, we have
\begin{eqnarray*}
\lefteqn{F_{b,b'}D(H)=2^{-(N-1)/2}\omega^{2b-N}\times}\\
&&\left\{\sqrt{N\choose b'}\ |1_y\rangle\langle S^y_{N-b,b}|
-i\sqrt{N\choose b}\ |0_y\rangle\langle S^y_{N-b',b'}|\right\}\\
&=&\omega^{2b-N-1}HF_{b,b'}.
\end{eqnarray*}
That is, each operator $F_{b,b'}$ yields a phase factor $\omega^{2b-N+1}$ when $H$ is applied. However, such factor does not change the operations of $F$ as a Kraus operator. For example, when applied to an arbitrary mixed state $\rho\in{\cal H}_B$, it yields
\[F_{b,b'}D(H)\rho D^\dagger(H)F_{b,b'}^\dagger=HF_{b,b'}\rho F_{b,b'}^\dagger H^\dagger.\]
By summing these over all combinations of $b,b'$ satisfying $b-b'\equiv1\ {\rm mod}\ 4$, we obtain Eq. (\ref{eq:F_invariance}).


\begin{thebibliography}{99}
\bibitem{BB84}
C. H. Bennett and G. Brassard,
in Proceeding of the IEEE International Conference on Computers, Systems, and Signal Processing,
Bangalore, India (IEEE, New York, 1984), pp.175-179.
\bibitem{Mayers}
D. Mayers, JACM {\bf 48}, 351 (2001).
\bibitem{LC98}
H.-K. Lo and H. F. Chau, Science {\bf 283}, 2050 (1999).
\bibitem{SP00}
P. W. Shor and J. Preskill, Phys. Rev. Lett. {\bf 85}, 441 (2000).
\bibitem{GL03}
D. Gottesman and H.-K. Lo, IEEE Trans. Inform. Theory {\bf 49}, 457 (2003), H. F. Chau, Phys. Rev. A {\bf 66}, 060302(R) (2002), K. S. Ranade and G. Alber, J. Phys. A: Math. Gen. {\bf 39}, 1701 (2006).
\bibitem{GLLP04}
D. Gottesman, H.-K. Lo, N. L\"utkenhaus, and J. Preskill, Quant. Inf. Comput. {\bf 5}, 325 (2004).
\bibitem{Koashi}
M. Koashi, arXiv:quant-ph/0609180v1.
\bibitem{information} B. Kraus, N. Gisin and R. Renner, Phys. Rev. Lett. {\bf 95}, 080501 (2005): R. Renner, N. Gisin and B. Kraus, Phys. Rev. A {\bf 72}, 012332 (2005).
\bibitem{decoy}
W. -Y. Hwang, Phys. Rev. Lett. {\bf 91}, 057901 (2003);
H.-K. Lo, X. Ma, and K. Chen, Phys. Rev. Lett. {\bf 94}, 230504 (2005);
X. -B. Wang, Phys. Rev. Lett. {\bf 94}, 230503 (2005).
\bibitem{Tamaki}
K. Tamaki, M. Koashi, and N. Imoto, Phys. Rev. Lett. {\bf 90}, 167904 (2003):
K. Tamaki and N. L\"utkenhaus, Phys. Rev. A {\bf 69}, 032316 (2004);
M. Koashi, Phys. Rev. Lett. 93, 120501 (2004);
K. Tamaki, N. L\"utkenhaus, M. Koashi, and J. Batuwantudawe,
arXiv:quant-ph/0607082v1.
\bibitem{FTL06}
K. Tamaki and H.-K. Lo, Phys. Rev. A. {\bf 73}, 010302(R) (2006); C.-H. F. Fung, K. Tamaki, H.-K. Lo, Phys. Rev. A {\bf 73}, 012337 (2006).
\bibitem{BBM92}
C. H. Bennett, G. Brassard, and N. D. Mermin, Phys. Rev. Lett. {\bf 68}, 557 (1992);
X. Ma, C.-H. F. Fung, H.-K. Lo, Phys. Rev. A {\bf 76}, 012307 (2007).
\bibitem{BLMS00}
G. Brassard, N. L\"utkenhaus, T. Mor, and B. C. Sanders, Phys. Rev. Lett. {\bf 85}, 1330 (2000);
N. L\"utkenhaus, Phy. Rev. A {\bf 61}, 052304 (2000).
\bibitem{Sakurai}
See, e.g., J. J. Sakurai, Modern Quantum Mechanics, Revised Edition (Addison-Wesley, Massachusetts, 1994).
\bibitem{BML08}
N. J. Beaudry, T. Moroder, N. L\"utkenhaus, arXiv:0804.3082v3.
\bibitem{KAYI08}
M. Koashi, Y. Adachi, T. Yamamoto, and N. Imoto, arXiv:0804.0891v1.
\end{thebibliography}
\end{document}